\documentclass[10pt,aps,prd,nofootinbib,reprint,groupedaddress]{revtex4-1}

\usepackage[utf8]{inputenc}
\usepackage{amsmath}
\usepackage{graphicx}
\usepackage{amssymb}

\usepackage{hyperref}
\hypersetup{colorlinks=true,allcolors=[rgb]{0.5,0.0,0.5}}

\newcommand{\orcid}[1]{\href{https://orcid.org/#1}{#1}}

\newcounter{RSQ}

\newcommand{\lsim}{\lesssim}
\newcommand{\gsim}{\gtrsim}
\newcommand{\eq}[1]{Eq.~(\ref{#1})}
\newcommand{\ord}[1]{\mathcal{O}{(#1)}}
\newcommand{\beq}{\begin{equation}}
\newcommand{\eeq}{\end{equation}}
\newcommand{\bea}{\begin{eqnarray}}
\newcommand{\eea}{\end{eqnarray}}

\newcommand{\thb}{\bar\theta}

\begin{document}

\title{Is the $\bar \theta$ parameter of QCD constant?}

\author{Hooman Davoudiasl}
\email{hooman@bnl.gov}
\thanks{\orcid{0000-0003-3484-911X}}

\author{Julia Gehrlein}
\email{jgehrlein@bnl.gov}
\thanks{\orcid{0000-0002-1235-0505}}

\author{Robert Szafron}
\email{rszafron@bnl.gov}
\thanks{\orcid{0000-0002-9640-6923}}

\affiliation{High Energy Theory Group, Physics Department, Brookhaven National Laboratory, Upton, NY 11973, USA}

\date{\today}

\begin{abstract}
Testing the cosmological variation of fundamental constants of Nature can provide valuable insights into new physics scenarios. While many such constraints have been derived for Standard Model coupling constants and masses, the $\thb$ parameter of QCD has not been
as extensively examined.  This letter discusses potentially promising paths to investigate the time dependence of the $\thb$ parameter. While laboratory searches for CP-violating signals of $\thb$ yield the most robust bounds on today's value of $\thb$, we show that CP-conserving effects provide constraints on the variation of $\thb$ over cosmological timescales. 
We find no evidence for a variation of $\thb$ that could have implied an ``iron-deficient"  Universe at higher redshifts.  By converting recent atomic clock constraints on a variation of constants, we infer $ d({\thb}^2)/dt \leq  6\times 10^{-15}\text{yr}^{-1}$, at 1-$\sigma$. 
Finally, we also sketch an axion model that results in a varying $\thb$ and could lead to excess diffuse gamma ray background, from decays of axions produced in high redshift supernova explosions.  
\end{abstract}

\date{\today}

\maketitle

\section{Introduction}

The modern understanding of particle physics compels us to treat any fundamental ``constant" of Nature as a possible parameter that could vary over time and space.  One of the earliest advocates of this view was Dirac \cite{Dirac:1937ti}, who attempted to explain why certain combinations of constants yield enormously large numbers.  According to his proposal, these numbers could be rationalized if one assumes that they start out as having natural values and then evolve over long cosmological times. 

Dirac's proposal is no longer the main motivation for considering variation of constants.  Nonetheless, it remains a possibility that values of various parameters in the Standard Model (SM) were, at very early times right after the Big Bang, different.  Masses of fermions, for example, are set by the Higgs field, after electroweak symmetry breaking (EWSB).  This also indicates that, if the  Universe started out very hot and dense, as is generally assumed, even the symmetries of the vacuum could have evolved, as would be the case for EWSB.  

At a more theoretical level, for example in the context of string theory, various constants of Nature are assumed to be set by the values of certain moduli, early on (see, {\it e.g.},  Ref.~\cite{Silverstein:2004id}).  However, one could imagine that these moduli may have continued to evolve over cosmological times leading to variations in the value of physical parameters, assumed to be constants (see  Ref.~\cite{Balasubramanian:2020lux} for a statistical interpretation; for early work in the context of extra-dimensional theories see   Ref.~\cite{Marciano:1983wy}).  This is the point of view we will adopt here. 
In particular, we will focus on variation of one parameter,  namely, the $\theta$ angle  of QCD, which is associated with the level of CP violation in strong interactions.

In Ref.~\cite{tHooft:1986ooh}, it was argued that $\theta$ cannot depend on space-time.  The gist of the argument is that $\theta$ parametrizes topological transformations corresponding to the winding number of QCD gauge configurations.  This notion will become ill defined if $\theta$ is a space-time-dependent field.  However, $\theta$ by itself is not a measurable quantity in  QCD.  Instead, a new quantity 
$\thb_q \equiv \theta + \arg[\det(M_q)]\,,
$
where $M_q$ is the quark mass matrix, is the effective parameter that would lead to CP-violating phenomena in QCD.  The smallness of $\thb_q\lsim 10^{-10}$, as implied by the upper bound on neutron electric dipole moment (EDM) $d_n < 1.8 \times 10^{-26}$~$e$ cm (90\% CL)  \cite{Abel:2020pzs, Baker:2006ts,ParticleDataGroup:2020ssz}, remains a conceptual puzzle and is often referred to as the ``strong CP problem."  

One of the most theoretically appealing resolutions of the above puzzle was proposed by Peccei and Quinn \cite{Peccei:1977hh,Peccei:1977ur}, by promoting $\thb_q$ to a field that relaxes to zero in the early Universe.  This is accomplished by introducing a global $U(1)$ symmetry, anomalous under QCD. Once the $U(1)$ is broken, a pseudo-Goldstone mode, called the axion \cite{Weinberg:1977ma,Wilczek:1977pj} and denoted by $a$ with decay constant $f_a$, would appear.  This field gets a mass, from a potential generated by nonperturbative QCD interactions. The axion has a minimum at $\langle a\rangle=-(f_a/\xi)\thb_q$, with $\xi$ an $\mathcal{O}(1)$ parameter, such that CP is conserved in strong interactions. 
The quantity of interest 
is now
$\thb$
\beq
\thb\equiv\thb_q-\xi(\langle a\rangle/f_a)~.
\eeq
In the following, we will assume that $\thb$ changes due to a change in the axion potential over time. 
For example, this change could come from  the effects of other light scalars (see appendix) such that $\thb$ is not constant in time. Alternatively, one could consider a change in $f_a$; a model for this has been proposed in Ref.~\cite{Allali:2022yvx}.
The $\thb$ parameter also runs in the SM, but the running starts at seven loops \cite{Ellis:1978hq}. This running allows us to introduce  a space-time dependence of $\thb$, through threshold corrections and new interactions with the background density (e.g. of dark matter), in analogy with the mechanisms discussed in \cite{Chacko:2002mf,Davoudiasl:2018ltz,Arakawa:2019dwr}.

In this {\it Letter}, we derive novel bounds on the cosmological time evolution of  $\thb$. Though there have been many past studies constraining the change of fundamental constants, and examining the  sensitivity of physical phenomena to the value of $\thb$ \cite{Hook:2017psm,Lee:2020tmi,2010AIPC.1269...21C,Ubaldi:2008nf}, this is the first study of the $\thb$ variation effects on proton-to-electron mass ratio over cosmological timescales, which allows us to set tighter bounds than previously obtained.  We will focus on the effect of changing $\thb$ on observables in atomic and nuclear physics. In models with an axion, additional constraints due to the presence of the axion field can also be derived, as discussed in the appendix.

\section{Signatures of varying $\mathbf{\thb}$ }
\label{sec:signatures}
The phenomenological consequences of $\thb$ are elusive, as they are typically not visible in perturbation theory. Nonetheless, the $\thb$ parameter is physical \cite{tHooft:1976rip,Callan:1976je,Jackiw:1976pf} and the theory is CP-conserving for $\thb=0,\pi$. We therefore think of $\thb$ as a continuous variable and distinguish between CP-conserving effects of varying $\thb$ and CP-violating effects. The most prominent CP-conserving effect of varying $\thb$ is the change in hadron masses. In fact, it can be shown that $\thb=\pi$ corresponds to a negative determinant of the quark mass matrix \cite{Crewther:1979pi}. We can construct a combination of the kaon and pion masses whose value is predicted by current algebra  \cite{Langacker:1978cf,weinberg:1977}
\begin{align}
r=\left( m_{K^0}^2-m_{K^\pm}^2- m_{\pi^0}^2+m_{\pi^\pm}^2\right)/m_\pi^2 = \frac{m_d \mp m_u}{m_d\pm m_u},
\end{align}
where the upper (lower) sign corresponds to $\thb=0$ ($\thb=\pi$).  Here, $m_u=2.16 $ MeV and $m_d=4.67$ MeV \cite{Zyla:2020zbs} are the up and the down quark masses, respectively.  The quantity $r$ is predicted to be less (greater) than $1$ for $\thb=0$  ($\thb=\pi$). We will consider small variations around $\thb=0$, since in Nature, the observed value of this ratio is less than $1$, and consequently, $\thb$ near zero is preferred today.

\subsection{CP-violating effects: EDM, atomic effects}

Before discussing the CP-conserving effects of non-zero $\thb$, we will briefly summarize the most prominent phenomenological consequences of CP violation induced by $\thb$. 
Nonzero $\thb$ implies a non-vanishing electric dipole moment (EDM) for hadrons. Intensive searches for neutron EDM (nEDM) so far have resulted in only an upper bound, which is the strongest experimental constraint: $\thb \lesssim 10^{-10}$ \cite{Abel:2020pzs, Baker:2006ts}. In  nEDM experiments, neutrons are placed in external electric and magnetic fields and one measures changes in their Larmor  precession \cite{Abel:2020pzs} (see  Refs.~\cite{ nEDM:2019qgk,Ito:2017ywc,Picker:2016ygp,Chanel:2018zga,Wurm:2019yfj,Serebrov:2017sqv,n2EDM:2021yah} for future proposals to improve the experimental sensitivity on the neutron EDM and Ref.~\cite{Omarov:2020kws} for  a  proposal using a proton storage ring to improve future experimental sensitivity to the proton EDM). 

For atomic systems, the EDM measurements are more challenging.  Schiff's theorem forbids effects linear in electron and proton EDMs. The theorem is valid in the nonrelativistic pointlike approximation; thus observable effects of EDMs are restricted  to higher order, relativistic effects, or effects related to the finite nucleus size. 

Experiments looking for EDMs in complex systems apply external fields and search for  tiny splittings of energy levels due to the nonzero EDM \cite{Sandars:1967zz}. A classic example of such searches is the measurement of the $^{199} \rm Hg$ EDM \cite{graner2016reduced}. If no source of CP violation other than $\thb$ is assumed, this measurement provides a strong bound $\thb \lesssim 1.5 \times 10^{-10}$, comparable to nEDM. However, theoretical interpolation of the experimental results is less clean due to numerous possible sources of CP violation in complex atomic systems and less precise computations of the relation between $\thb$ and  $^{199} \rm Hg$ EDM \cite{Engel:2013lsa,Mereghetti:2018oxv}.

We see that CP violation effects of $\thb$ are nontrivial to observe and require precise control over external fields. While such conditions can be realized in  terrestrial experiments, it is not feasible to search for EDMs over astronomical distances.  We therefore turn to the CP-conserving effects of $\thb$, which though less sensitive, are much easier to observe over astronomical distances. 

For a recent review about EDMs see Ref.~\cite{Pospelov:2005pr}; for reviews about using atoms to constrain new physics see Refs.~\cite{Ginges:2003qt,Safronova:2017xyt}.

\subsection{CP-conserving effects: Hadron masses, molecular effects}

The value of $\thb$ affects
various hadronic properties like the proton and neutron masses but also binding energies  of nuclei. Of particular interest to obtain constraints on $\thb$ is the dependence of the pion mass on $\thb$ as this affects the nucleon masses, the neutron-proton mass difference, and the neutron decay width, which play a role in Big Bang Nucleosynthesis (BBN).

The  leading order $\thb$-dependence  of the pion mass  in the  two-flavor  approximation  is \cite{Leutwyler:1992yt,Brower:2003yx}
\begin{align}
    m_\pi^2(\thb)=m_\pi^2\cos(\thb/2)\sqrt{1+\epsilon^2\tan^2(\thb/2)}\,,
\end{align}
where the pion mass $m_\pi=m_\pi(\thb=0)=139.57 $ MeV and $\epsilon=(m_d-m_u)/(m_d+m_u)\approx0.37$  quantifies the departure from the
isospin symmetric limit  $\epsilon =0$.
With this expression,  the nucleon mass  in the $\epsilon \to 0$ limit is given as \cite{Brower:2003yx}
\begin{align}
    m_N(\thb)=m_0-4c_1 m_\pi^2(\thb)-\frac{3 g_A^2 m_\pi^3(\thb)}{32 \pi f_\pi^2}\,,
    \label{eq:nucleonmass}
\end{align}
with the nucleon mass in the chiral limit  $m_0=869.5$ MeV \cite{Hoferichter:2015hva}, $g_A = 1.27$ is the axial-vector coupling constant,
$f_\pi = 92.2$ MeV is the pion decay constant, and $c_1 = -1.1~\text{ GeV}^{-1}$ \cite{Hoferichter:2015tha} is a low-energy constant  from the second-order chiral pion-nucleon Lagrangian (see  Ref.~\cite{Bernard:1995dp} for a review on this topic).
The neutron mass is well approximated by Eq.~(\ref{eq:nucleonmass}); for the proton mass one needs to include the $\thb$-dependence of the QCD contribution to the neutron-proton mass difference \cite{Lee:2020tmi,Vonk}
\begin{align}
   ( m_n-m_p)^{QCD}(\thb)\simeq4 c_5 B_0\frac{m_\pi^2}{m_\pi(\thb)^2} (m_u -m_d)\;,
   \label{eq:mn-mp}
\end{align}
with $B_0=m_\pi^2/(m_u+m_d)$ and $c_5=(-0.074 \pm 0.006) ~\text{GeV}^{-1}$.
The pion mass affects the strength of the nuclear force.
However, the effects of $\thb$ on a multinucleon system are difficult to quantify; we only have  a qualitative notion that with increasing $\thb$ the nuclear binding energy will increase and  the relative importance of the
Coulomb interaction will decrease \cite{Lee:2020tmi}. An increase in the binding energy mimics a lighter nucleus, and so does a decrease in the nucleon mass.
Therefore, to obtain   a conservative lower bound on the effect of a variation of $\thb$ in systems with many nucleons, we will focus on the effect due to the shift in the nucleon mass only.
This allows us to make use of the very powerful data from spectroscopic measurements of molecular transitions at various redshifts. 
These observations have been used to constrain the ratio of proton-to-electron mass, $\mu \equiv m_p/m_e$, during the evolution of the Universe, as it affects atomic transitions \cite{Ubachs:2015fro,Martins:2017yxk}. Under the assumption that the effect of the nucleon mass change\footnote{The change in the electron mass is negligible  as the electron does not couple directly to QCD.} is the dominant effect of $\thb$, we can obtain a conservative bound on $\Delta{\thb}$ from changes of $\mu$. We constrain ourselves to using  few-nucleon systems like  H$_2$ and HD (hydrogen deuteride) in order to obtain the most reliable bounds. In the appendix, we provide details on our treatment of di-atomic molecules.
By using the ratio of the expression in  \eq{eq:nucleonmass} for free $\thb$ over that  evaluated at $\thb=0$,
we arrive at the relation between $\mu$ and $\Delta(\thb^2)$ as
\begin{align}
    \Delta(\thb^2) \approx-1.4\times10^{2}~ \frac{\Delta\mu}{\mu}\;.
    \label{eq:thetavar}
\end{align}
Notice that a change in $\thb$ can only decrease the nucleon masses and therefore $\Delta\mu/\mu$.
Even though the  deuteron is a multi-nucleon system where instead of the one-pion exchange the exchange of the $\omega,~\rho,~\sigma$ mesons determines the binding energy \cite{Lee:2020tmi},  we find that  the numerical dependence of a change in the deuteron mass on a variation in $\thb$ is comparable and even slightly smaller than for a change in $\mu$.
Therefore, including HD data (which, in fact represents a rather small fraction of the total dataset) leads to conservative bounds on $\Delta (\thb^2$).
We show the constraints on $\Delta(\thb^2)$ from H$_2$ and HD observations in Table~\ref{tab:theta_variation} for various values of the redshift.

\begin{figure}
\includegraphics[width=0.45\textwidth]{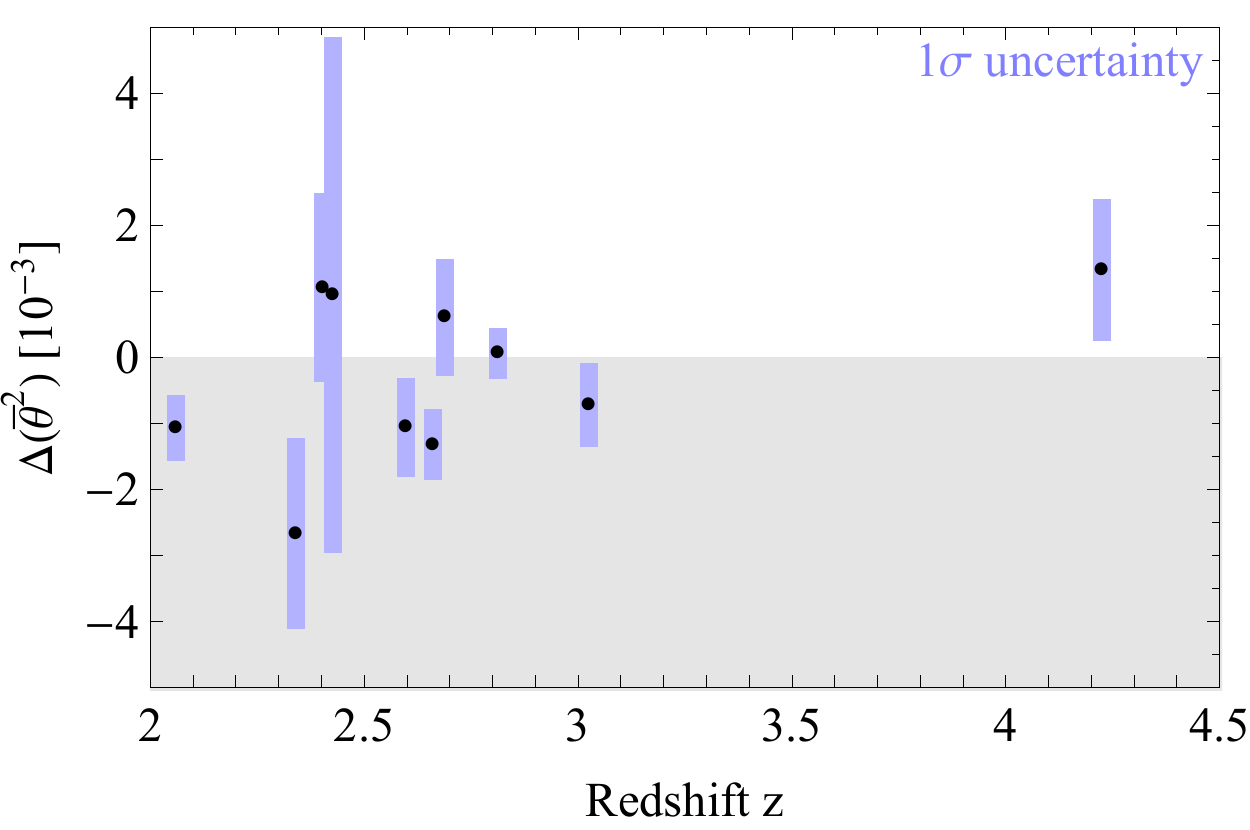}

\caption{Constraints on  $\Delta(\thb^2)$ derived from bounds on $\Delta \mu/\mu$ in  tab.~\ref{tab:theta_variation} across a range of redshifts $z$. The two data points at $z=2.659,~2.66$ have been combined and averaged.
The best fit points  are shown as black circles and  the $1\sigma$ uncertainties are shown in blue. Some data points from tab.~\ref{tab:theta_variation} lead to negative $\Delta (\thb^2)$ and therefore to unphysical values of $\thb$ which we show in the gray region.   }
\label{fig:thetavar}
\end{figure}

Using the upper limit on $\thb$ from neutron EDM experiments on Earth \cite{Abel:2020pzs, Baker:2006ts} 
we can constrain $\thb$ during the evolution the Universe.  In Fig.~\ref{fig:thetavar}, we plot the best fit values of $\Delta (\thb^2)$ and their $1\sigma$ errors.  Since today's value of $\thb$ is bounded to be smaller than $\sim 10^{-10}$, the corresponding bound on $\thb$ in this figure can, to a very approximation, be obtained from $\sqrt{\Delta (\thb^2)}$. Some of the results for $\Delta (\thb^2)$ in Table~\ref{tab:theta_variation}  are negative, which leads to unphysical values of $\thb$. These correspond to best fit values in the gray region of the figure.  We deduce from Fig.~\ref{fig:thetavar} that 
the data used in our study, taken all together, offer no evidence for any time variation in $\thb$ and strongly favor $\Delta (\thb^2)=0$.

Our results are compatible and even stronger than direct constraints on the value of $\thb$ in the early Universe coming from the $^4\text{He}$ mass fraction  at BBN  \cite{Lee:2020tmi}, stellar dynamics \cite{2010AIPC.1269...21C,Lee:2020tmi, Hook:2017psm,Ubaldi:2008nf},
 X-ray emissions  from the surroundings of compact stellar objects \cite{Hook:2017psm}, and the measured value of the proton and neutron mass today.
    
 In fact, a change in $\thb$ affects the abundance of   $^{56}\text{Fe}$ and $^{56}\text{Co}$ in the Universe
as $^{56}\text{Co}$ would be the most tightly bound nucleus instead of  $^{56}\text{Fe}$ for large $\thb$.  The increase in the neutron-proton mass difference for large $\thb$ \cite{Ubaldi:2008nf} leads to $^{56}\text{Fe}$ being heavier than $^{56}\text{Co}$ by $\sim 5$ MeV \cite{Hook:2017psm}, and $^{56}\text{Fe}$ produced in the stars
would have decayed to $^{56}\text{Co}$. Therefore, large $\thb$ leads to an ``iron-deficient Universe". From this effect, one can also derive an upper limit of $\thb\lesssim \mathcal{O}(1)$ \cite{Hook:2017psm} making use of the observation of the Fe K$\alpha$ line
around white dwarfs and neutron stars 
combined with the non-observation of a  $^{56}\text{Co}$ line. 
Variation of $\thb$ could also affect the shape of the light curve of type Ia supernovae, through its effect on the mass of  $^{56}\text{Co}$ whose radioactive decay is 
the dominant heating
source for the supernova remnant \cite{Scherrer:1992na}.

\begin{table}
\centering
\caption{Constraints on the variation of $\thb^2$ for various redshifts making use of different  systems. We reinterpret the constraints on $\Delta \mu/\mu$ coming from methods involving H$_2$ and HD using \eq{eq:thetavar}. The errors indicate $1\sigma$ uncertainties. }
\begin{tabular}{c|c|c|c}
$z$&$ \Delta(\thb^2)$ $(\times 10^{3})$&Object&Ref\\\hline
 2.059&$-1.06\pm0.49$&J2123-005 &\cite{vanWeerdenburg:2011ru,Malec:2010xv,Ubachs:2015fro}\\
 2.34&$-2.66\pm1.44$&Q1232+082&\cite{Dapra:2016dqh}\\\
2.402 &$1.06\pm1.44 $ &HE0027-1836 &\cite{Rahmani:2013dia}\\
2.426&$0.95\pm 3.89$&Q2348-011 &\cite{Bagdonaite:2011ab}\\
2.597&$-1.05\pm0.74$&Q0405-443&\cite{Ivanchik:2005ws,Reinhold:2006zn,King:2008ud,Thompson:2009tn,Ubachs:2015fro}\\
2.659 &$-1.04\pm 0.93$&J0643-504 &\cite{AlbornozVasquez:2013wfw}\\
2.66&$-1.44\pm0.64$&B0642-5038&\cite{AlbornozVasquez:2013wfw,Dapra:2015yva,Ubachs:2015fro}\\
2.688 &$0.61\pm0.88$&J1237+0648&\cite{Ubachs:2018wan}\\
2.811 &$0.07\pm0.38$&Q0528-250&\cite{Cowie:1995sz,Ledoux:2003mv,King:2011km,King:2008ud,Ubachs:2015fro}\\
3.025&$-0.71\pm 0.63$&Q0347-383&\cite{King:2008ud,Thompson:2009tn,Wendt:2010qe,Wendt:2012ea,Ubachs:2015fro}\\
4.224&$1.33\pm 1.06$ &J1443+2724 &\cite{Bagdonaite:2015kga}\\
\end{tabular}
\label{tab:theta_variation}
\end{table}

Local time variation of physical constants is currently constrained very precisely by atomic clocks \cite{Safronova:2017xyt}.  In particular, recent measurements have yielded $\dot\mu/\mu = -8 \pm 36 \times 10^{-18}~\text{yr}^{-1}$ \cite{Lange:2020cul}. Using this result, and employing  Eq.~(\ref{eq:thetavar}) as a conservative bound on the conversion factor between the  $\dot\mu/\mu $ and the $d({\thb}^2)/dt$ for complex nuclei (such as $^{171} \text{Yb}^+$ used in Ref.~\cite{Lange:2020cul}), we obtain the upper bound on the local time variation of $\thb^2$ 
\begin{align}
    \frac{d({\thb}^2)}{dt}\leq  6\times 10^{-15}\text{yr}^{-1}\;,
\end{align}
where the bound should be interpreted as corresponding to $1 \sigma$ level. Without a specific model for the time dependence of $\thb$, one cannot infer a bound on $\thb$ at earlier epochs.


\section{Summary $\mathbf{\&}$ Conclusions}
\label{sec:conclusions}
Testing the constancy of fundamental constants of the SM can provide valuable insights into physics beyond the SM. 
In this {\it Letter} we established for the first time constraints on the variation of the $\thb$ parameter of QCD over cosmological timescale from data on molecular transitions at different redshifts. As $\thb$ affects hadronic properties a change in $\thb$ translates to a change in the proton-to-electron mass ratio that has been constrained with various observations. Making use of the observations involving H$_2$ and HD molecules we find that generally $\thb\lesssim 0.1$ for redshifts $z\sim 2-4$. By converting  atomic clock constraints on the local variation of constants, we infer $d({\thb}^2)/dt\leq  6\times 10^{-15}\text{yr}^{-1}$.

Our results constrain models that predict a change in $\thb$ at late times while various direct limits on $\thb$, including bounds  from early Universe physics, lead to weaker constraints of $\thb \lesssim 1$ 
\cite{Lee:2020tmi,2010AIPC.1269...21C,Lee:2020tmi, Hook:2017psm,Ubaldi:2008nf}.
To further constrain variations of $\thb$ in the future one could make use of the effects of  $\thb$ on nuclear properties. For example, for varying $\thb$  we also expect long-lived
isotopes to become short-lived and hence rarer than observed, or as the 
phase space changes,  some decays might become forbidden. These studies require complicated nuclear calculations, which is beyond the scope of this letter.

Other effects of varying $\thb$ are more model dependent, for example signatures related to the change in the axion mass (see  appendix for a discussion). If the change in $\thb$ is due to a shift in the axion potential larger $\thb$ for high $z$ corresponds to heavier axions in the early Universe. This could lead to the prediction of gamma rays from the  decay of axions produced in supernovae or neutron stars.

Disentangling a change in $\thb$ from a change in other quantities like $\alpha,~\mu$, and  $\alpha_s$ is challenging; however, a few avenues exist. 
Varying $\alpha$ leads to a change in the fine-structure doublets whereas a change in $\mu$ can be measured by comparing molecular
hydrogen vibrational and rotational modes \cite{Martins:2017yxk}. Furthermore,
the quantities $\alpha,~\mu$, and $\alpha_s$ are not related to parity violation, whereas
$\thb$ is generally assumed to be a measure of parity violation in QCD. Therefore, if $\thb$ was larger at high $z$, one could expect electric and magnetic multipole nuclear transitions to mix. This could affect the relative intensity of various spectral lines. Its observation could be used to distinguish $\thb$ variation from a change in $\alpha_s$ or $\mu$, which would have a homogeneous effect on the spectrum. This effect is solely due to the parity-violating effects of $\thb$. However, the $\thb$ induced parity non-conserving forces would be spin-dependent and, consequently, subleading to the dominant parity-conserving binding effects. 

In conclusion, any discovery that establishes time-dependence of fundamental parameters would have revolutionary implications for our understanding of the Universe. In this context, searches for cosmological-scale space-time-dependence of the $\thb$ parameter are well motivated and it is essential to explore new avenues.

\begin{acknowledgments}
We thank Bob Scherrer for pointing out the relation between the lifetime of $^{56}\text{Co}$ and the light curve of supernovae type Ia.
 The authors acknowledge support by the United States Department of Energy under Grant Contract No.~DE-SC0012704.
\end{acknowledgments}

\appendix \section{Model motivation }
\label{sec:model}
In this appendix we present an example model  which leads to a time-dependent  $\thb$ and study aspects of its phenomenology.
\subsection{A model for changing $\thb$ }
There could be various physical effects that can shift the minimum of an axion potential away from $\langle a\rangle=-(f_a/\xi)\thb$.  These include new QCD CP-violating operators \cite{Pospelov:2005pr} and explicit Peccei-Quinn $U(1)$ breaking effects mediated by gravity   \cite{Holman:1992us,Kamionkowski:1992mf,Barr:1992qq,Ghigna:1992iv,Kallosh:1995hi}, since gravitational effects, such as microscopic black holes, are generally expected to violate global symmetries.  As an example of the latter, let us consider a dim-5 operator 
\beq
O_{PQb} = \kappa \frac{\varphi + \varphi^*}{M_P}|\varphi|^4\,,
\label{OPQb}
\eeq
where $\varphi$ has PQ charge $+1$ and whose vev $f_a$ breaks the $U(1)$ spontaneously; $\kappa$ is a numerical coefficient determined by UV physics and $M_P\approx 1.2\times 10^{19}$~GeV is the Planck mass.  This operator explicitly breaks the PQ symmetry and hence would induce a correction to the axion mass given by
\beq
\delta m_a^2 \sim \kappa \frac{f_a^3}{M_P}.
\label{delma2}
\eeq

Note that for the axion to solve the strong CP problem, its potential must be dominated by non-perturbative QCD effects, which induce an axion mass
\beq
m_a^2 \sim \frac{m_\pi^2 f_\pi^2}{f_a^2}\,,
\label{ma2}
\eeq
where $m_\pi$ and $f_\pi$ are the pion mass and decay constant, respectively.  One can show that the above shift in axion mass from \eq{delma2} would imply a shift in $\thb$ given by
\beq
\delta \thb \sim \frac{\delta m_a^2}{m_a^2}\sim \frac{\kappa f_a^5}{m_\pi^2 f_\pi^2 M_P}.
\label{delthb}
\eeq
This shift in the apparent value of $\thb$ is a result of the shift in the minimum of the the axion potential, such that $ \thb_q-\xi\langle a\rangle/f_a $ no  longer vanishes.

Astrophysical constraints imply that $f_a \gsim 10^8$~GeV \cite{Zyla:2020zbs}.  The above shift must satisfy  $\delta \thb \lsim 10^{-10}$ in order to satisfy the current bounds on $\thb$ and, hence, one needs $\kappa \lsim 10^{-35}$.  Consequently, if $\kappa$ can vary by $\ord{10^9}$ over cosmological times and distances, one could have $\thb\sim 1$ across such space-time scales.  The required tiny effect from such operators poses a problem for axion models.  Nonetheless, one can imagine that the smallness of $\kappa$ could be related to certain UV scale instantons \cite{Hui:2016ltb} that are needed to violate PQ charges conservation.  Generally, instanton effects are suppresses by $\sim e^{-2\pi/\alpha_g}$, where $\alpha_g$ denotes some coupling constant; to get $\kappa \lsim 10^{-35}$ ``today" we need $\alpha_g\lsim 1/13$.  The above analysis suggest that a change in  $\alpha_g$ by $\sim 30\%$ could achieve the possible variation in $\thb$.  This may be a result of a modest  shift in some modulus $\phi$, $\delta \phi$, that couples to the gauge field strength $G_{(g)\mu\nu}$ associated with $\alpha_g$ through
\beq
\frac{\phi}{M_P}\, G_{(g)\mu\nu}G_{(g)}^{\mu\nu}\,,
\label{phiGG}
\eeq
where $\delta \phi/M_P\sim  \delta (4\pi\alpha_g)^{-1}$.

\subsection{Phenomenology of a shift in the axion potential}

An interesting phenomenological consequence of interpreting $\delta \thb$ as a shift in the axion potential is that axions can become much heavier at $z\gsim 1$ and potentially decay over cosmological time scales.  Given that our preceding analysis of cosmological constraints allows $\thb \sim {\rm few} \times 10^{-2}$ for $2 \lsim z\lsim 4$, \eq{delthb} suggests that the axion mass could have been about 4 orders of magnitude larger than today's value required by locally measured observables.  Let us take $f_a\sim 3\times 10^8$~GeV as an example, near the edge of the allowed parameter space.  Since $f_a$ is set by UV physics typically unrelated to the dynamics that sets the axion potential, we will take it to be constant.  Hence, from \eq{ma2}, today's value of axion mass is given by $m_a(z=0)\sim 0.03$~eV, suggesting that at earlier epochs it could have been as large as $m_a(z\gsim 2) \sim 0.1$~keV.

The above high-redshift value of the axion mass is still small enough that $a$ could have been emitted by core collapse supernovae, whose temperatures are typically tens of MeV \cite{Zyla:2020zbs}, corresponding to energies $E_a$ of $\ord{100~{\rm MeV}}$.  Hence,  axions of mass $\sim 0.1$~keV will be emitted by the supernova with a boost $E_a/m_a\sim 10^6$, for our choice of parameters.  One can then estimate the decay length $l_a$ of such an axion 
\beq
l_a \sim \frac{E_a}{m_a\Gamma_a}\,,
\label{la}
\eeq
where the axion's decay width is given by \cite{Zyla:2020zbs}
\beq
\Gamma_a = \frac{g_{a\gamma\gamma}^2 \, m_a^3}{64\pi}\,,
\label{Gammaa}
\eeq
and the axion-photon coupling $g_{a\gamma\gamma}\sim \alpha/(2\pi f_a)$.  We hence find $l_a \sim 3\times 10^{19}$~yr.  However, the redshift interval corresponding to the larger axion mass corresponds to $\sim {\rm few} \times 10^9$~yr.  Thus, we roughly expect that only a fraction $\ord{10^{-10}}$ of the supernova axions that were emitted then would have decayed before their masses decreased at later epochs.

The rate of core collapse supernovae (CCSN) is approximately given by $R_{cc}\sim 1.25 \times 10^{-4}$~Mpc$^{-3}$~yr$^{-1}$ and grows as $10^z$ until $z\approx 1$, after which it is roughly constant \cite{Horiuchi:2008jz,Raffelt:2011ft}.
With this information and based on the formalism presented in Ref.~\cite{Raffelt:2011ft}, we estimate that the contribution of the CCSN to the axion diffuse background between $z\approx2-4$ could have been as large of $\sim 1/4$ of the total axion flux from Supernovae.  The decay $a\to \gamma\gamma$ would then lead to a photon flux roughly at the level of $\ord{10^{-11}}$ of the axion flux, characterized by half the average energy of the emitted axions $E_\gamma \sim 10$~MeV, accounting for the assumed typical redshift.  Using the results of Ref.~\cite{Raffelt:2011ft}, we hence deduce that the contribution to the diffuse gamma ray flux at $10$~MeV would be at the level of $\sim 10^{-10}$~cm$^{-2}$~s$^{-1}$, which is well below the measured value $\sim 10^{-3}$~cm$^{-2}$~s$^{-1}$ for such energies \cite{Fornasa:2015qua}.  However, we note that since 
$l_a\propto f_a^6$, a mildly smaller value of $f_a$ can lead to a large enhancement of the possible gamma ray flux, by allowing significantly more axions to decay in the $z\sim 2-4$ era, this would provide potential constraints on the model.     
For related studies, see Refs.~\cite{Calore:2020tjw,Calore:2021hhn}.

\section{Details of treatment of di-atomic molecules}

To understand the dependence of the di-atomic molecule energy levels on the nucleus mass, we first consider the  Born-Oppenheimer approximation  \cite{born1927quantentheorie} for the energy levels
\begin{align}
    E = E_{\rm el} + \chi^2 E_{\rm vib} +  \chi^4 E_{\rm rot} + \ldots \;,
\end{align}
where the expansion parameter $\chi = (m_e/M)^{1/4}$ is inversely proportional to the reduced  nuclear mass $M$. The scaling of various terms can be easily understood from simple quantum mechanical considerations \cite{RevModPhys.3.280}. $E_{\rm el}$  is the electronic energy, which in the leading approximation does not depend on the mass of the nuclei forming the molecule. The next term, $E_{\rm vib}$ represents the vibrational energy. This contribution can be modeled by a quantum mechanical oscillator, whose energy levels are inversely proportional to the square root of the vibrating mass \cite{Born:1925mph}. Finally,
$E_{\rm rot}$ is the rotational energy. Energy levels of a quantum rotor are inversely proportional to the moment of inertia \cite{PhysRev.28.318}, which in turn depends linearly on the mass. 
In practice, a more accurate description is obtained using the Dunham expansion \cite{PhysRev.41.721}
\begin{align}
    E(\nu,J) = \sum_{k,l} Y_{k,l} \; (\nu +1/2)^k (J(J+1))^l \;,
\end{align}
where $J$ is a rotational quantum number and $\nu$ is a vibrational quantum number.  Constants $Y_{k,l}$ are known as the Dunham parameters and can be calculated for small oscillations. In particular, $Y_{0,0}=E_{\rm el}$ represents electronic energy. 
Importantly for us, $Y_{k,l}$ are homogeneous functions of the reduced nuclear mass $M$ in the leading approximation \cite{PhysRev.25.119}. This mass dependence can be exploited to obtain experimental bounds on proton (nucleus) mass variation \cite{thompson1975determination} (see also Refs.~\cite{Varshalovich:1993sb,Karshenboim:2000rg}), which we in turn interpret as the effect of a $\thb$ variation.

For more complex molecules than H$_2$ one is measuring the ratio of the effective nuclear mass to the electron mass unlike it is the case for molecular hydrogen. In general, the relative variation of this ratio will be equal to that of $\mu$ if  there are no composition-dependent
forces—in other words, this means if  the proton mass, neutron mass, and biding energy change in an identical way \cite{Martins:2017yxk}.
For HD we only consider the change in the proton and neutron mass, neglecting the change in the binding energy which leads to a  conservative bound as the binding energy only increases with increasing $\thb$.
\bibliography{main}

\end{document}